\documentstyle[aps,prl,preprint,floats,epsfig]{revtex}

\begin{document}

\preprint{\tighten\vbox{\hbox{\hfil CLNS 99/1655}
                        \hbox{\hfil CLEO 99-21}
}}

% Comment out to get double spacing.
%\tighten

%\title{

%\rightline{\normalsize CLNS 99/NNNN, CLEO 99-N, DRAFT 99-38A}
%
%\rightline{\normalsize Author: Vladimir Savinov (savinov@mail.lns.cornell.edu)}
%
%\rightline{\normalsize Committee: Nari Mistry (chair, nbm@mail.lns.cornell.edu), Jim Fast, Yongsheng Gao}
%
%\rightline{\normalsize This draft released at 3:00 pm PST, 12/13/99}
%
%\vspace{1.0cm}

%\noindent Dear CLEO Collaborator, 
%
%\vspace{2mm}
%
%\noindent The document in front of you is not in its final shape. 
%The author and the members of the committee 
%will continue to work on improvements of the text 
%after the internal CLEO submission deadline 
%for the December'99 Collaboration meeting vote. 
%Your comments and suggestions are highly appreciated. 
%Please send these by e-mail to the author and the chair. 
%We plan to submit this paper to PRL. 
%
%\vspace{2mm}
%
%~~~~~~~~~~~~~~~~~~~~~~~~~~~~~~~~Thank you (also acting on behalf of the Committee), 
%
%\vspace{1mm}
%
%~~~~~~~~~~~~~~~~~~~~~~~~~~~~~~~~~~~~~~~~~~~~~~~~~~~~~~~~~~~~~~~~~~~~~~~~~~~~~~~~~~~~~~~~~~~~~~~~~~~~~~Vladimir Savinov
%
%\noindent p.s. in case you find it useful my fax is (650) 926-4001
%

%\vspace{3.0cm}
%
%\centerline{\large\bf Search for the Decay $\bar{B^0} \to D^{*0}\gamma$}
%
%\vspace{0.5cm}
%
%\centerline{\normalsize(\today)}
%
%\vspace{0.5cm}
%
%\centerline{\normalsize CLEO Collaboration}
%
%}  

% Insert author list file here

\title{Search for the Decay $\bar{B^0} \to D^{*0}\gamma$}

\author{(CLEO Collaboration)}

\date{\today}

\maketitle
\tighten

\begin{abstract} 
% Insert abstract here.

We report results of a search for the rare radiative decay 
$\bar{B^0} \to D^{*0} \gamma$. 
Using $9.66 \times 10^6$ $B\bar{B}$ meson pairs 
collected with the CLEO detector at the 
Cornell Electron Storage Ring, we set 
an upper limit on the branching ratio for this decay 
of $5.0 \times 10^{-5}$ at 90\% CL. 
This provides evidence 
that anomalous enhancement 
%which could have overcome 
%the short-distance color suppression 
is absent in $W$-exchange processes 
and that weak radiative $B$ decays 
are dominated by the short-distance 
$b \to s\gamma$ mechanism in the Standard Model. 

\end{abstract}

\newpage 

{
\renewcommand{\thefootnote}{\fnsymbol{footnote}}

\begin{center}
M.~Artuso,$^{1}$ R.~Ayad,$^{1}$ C.~Boulahouache,$^{1}$
K.~Bukin,$^{1}$ E.~Dambasuren,$^{1}$ S.~Karamov,$^{1}$
S.~Kopp,$^{1}$ G.~Majumder,$^{1}$ G.~C.~Moneti,$^{1}$
R.~Mountain,$^{1}$
% ...
\end{center}
 
\small
\begin{center}
$^{1}${Syracuse University, Syracuse, New York 13244}\\
% ...
\end{center}

\setcounter{footnote}{0}
}

%\pacs{PACS numbers: }

\newpage

% Begin main body of text.

%++++++++++++++++++++++++++++++++++++++++++++++++++++++++++++++++++++++++++++++++++++++++++++

In recent years exclusive\cite{CLEO:discovery_bsg_exclusive} 
and inclusive\cite{CLEO:discovery_bsg_inclusive} 
$b \to s\gamma$ transitions 
were discovered by CLEO. 
These observations confirmed 
the existence of effective 
flavor changing neutral current 
processes in the Standard Model (SM) 
and stirred significant theoretical interest 
by opening new avenues to search 
for new physical phenomena\cite{new_physics}. 

One of the essential ingredients 
of the inclusive $b \to s\gamma$ 
measurement by CLEO was the assumption that 
flavor annihilation and $W$-exchange radiative transitions,  
represented by decays such as 
$\bar{B^0} \to D^{*0}\gamma$,  
are strongly suppressed. 
If this were not so, 
these decays could represent a 
serious experimental background to 
the inclusive photon spectrum used 
to deduce the $b \to s\gamma$ rate. 
The primary goal of the study 
presented in this Letter 
is to establish experimentally 
whether $W$-exchange (flavor annihilation) 
processes are indeed strongly suppressed in $B$ decays. 

\begin{figure}[ht]
\centerline{
%clns	\psfig{figure=prl0.eps,height=1.50in,width=3.20in}
	\psfig{figure=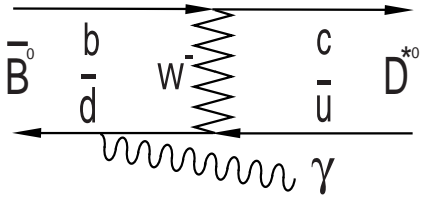,height=1.10in,width=2.20in}
%draft	\psfig{figure=prl0.eps,height=1.50in,width=3.20in}
           }
\caption{
One of lowest order Feynman diagrams for 
the decay $\bar{B^0} \to D^{*0} \gamma$ in 
the Standard Model. 
}
\label{diagram}
\end{figure}

We search for the decay 
$\bar{B^0} \to D^{*0}\gamma$ 
(and its charge conjugate state). 
In the SM framework this decay proceeds 
via $W$-exchange 
between $b$ and $\bar{d}$ quarks (Fig.~\ref{diagram}). 
Naively, this transition is suppressed 
by helicity effects and 
Quantum Chromodynamic (QCD) color corrections 
to the weak vertex. Two theoretical mechanisms 
to overcome this suppression have been proposed 
in the past. One mechanism has to do with the emission 
of gluons from the initial state quark\cite{theory_1} while 
the other\cite{theory_2} assumes a large $q\bar{q}g$ (or color octet) 
component in the $B$ meson wave function.
Whether either mechanism could significantly 
enhance the rate is debatable\cite{theory_3}. 
Theoretical estimates which take 
gluon emission into account 
predict a $\bar{B^0} \to D^{*0}\gamma$ branching 
fraction of the order of $10^{-6}$\cite{theory_3,theory_4,theory_5}. 
Though the numerical estimates of the rate for the 
color octet hypothesis are not yet available, 
it is expected that the rate could be enhanced by a 
factor of approximately ten 
which is a typical color suppression factor. 
So far the presence of a possible enhancement 
in the decay $\bar{B^0} \to D^{*0}\gamma$ 
has not been tested experimentally. 

On the other hand, if QCD suppression 
is present in the decay $\bar{B^0} \to D^{*0}\gamma$, 
eventually we would like to measure the strength of this suppression. 
Theoretical predictions for the studied decay 
have large uncertainties, therefore,  
a precise knowledge of the branching fraction 
would allow the QCD radiative 
corrections to be quantified 
more reliably. 
Knowledge of these corrections 
becomes increasingly important as 
theorists suggest new ways to 
constrain the SM parameters 
using hadronic $B$ decays. 
This makes the decay $\bar{B^0} \to D^{*0}\gamma$ 
an interesting process to study even if 
QCD suppression is present. 

The data analyzed in this study were collected at the 
Cornell Electron Storage Ring (CESR) 
with the CLEO detector. 
The results are based on $9.66 \times 10^6$ 
$B\bar{B}$ meson pairs, corresponding to 
an integrated $e^+e^-$ luminosity of 
$9.2 fb^{-1}$ collected at the $\Upsilon(4{\rm S})$ energy 
of 10.58 GeV. 
To optimize most of our selection criteria, 
we also employed $4.6 fb^{-1}$ of 
$e^+e^- \to q\bar{q}$ ($q = u,d,s,c$) 
annihilation data (``continuum'') 
collected approximately 60 MeV below the $\Upsilon(4{\rm S})$ energy. 
Our data sample was recorded with two configurations 
of the CLEO detector. 
The first third of the data were recorded with 
the CLEO II detector\cite{CLEOII:description} 
which consisted of three cylindrical drift chambers 
placed in an axial solenoidal magnetic field of 1.5T, 
a CsI(Tl)-crystal electromagnetic calorimeter, 
a time-of-flight plastic scintillator system 
and a muon system (proportional counters embedded 
at various depths in the steel absorber). 
Two thirds of the data were taken with 
the CLEO II.V configuration of the detector 
where the innermost drift chamber 
was replaced by a silicon vertex detector\cite{CLEOII.V:description} 
and the argon-ethane gas of the main drift chamber 
was changed to a helium-propane mixture. 
%The former upgrade improved 
%the ability of the detector 
%to reconstruct the decay lengths 
%of (relatively) long-lived particles. 
This upgrade led to improved resolutions in momentum 
and specific ionization energy loss 
($dE/dx$). 
The response of the detector is modeled with 
a GEANT-based\cite{GEANT} Monte Carlo simulation program. 
The data and simulated samples are processed 
by the same event reconstruction program. 
Whenever possible the efficiencies are either 
calibrated or corrected for the difference 
between simulated and actual detector responses 
using direct measurements from independent data. 
%The components of the detector most relevant 
%to this analysis are the tracking devices 
%and electromagnetic calorimeter. 

We search for 
$\bar{B^0} \to D^{*0}\gamma$ 
candidates among events where 
a photon with energy 
greater than 1.5 GeV 
is accompanied by a fully 
reconstructed $D^{*0}$ meson. 
The $D^{*0}$ mesons 
are reconstructed in their decays 
to $D^0 \pi^0$ and $D^0 \gamma$ 
with the $D^0$ mesons decaying 
to $K^-\pi^+$, 
$K^-\pi^+\pi^0$ 
or 
$K^-\pi^+\pi^-\pi^+$. 
These reconstructed channels 
comprise 25\% of the 
product branching fraction 
for the $D^{*0}$ and $D^0$ decays. 
Multiple entries are assigned 
a weight inversely proportional to the 
number of candidates identified in the event. 
As we apply selection criteria, 
the reweighting is performed appropriately. 
%Therefore each event always 
%enters the analysis with unit weight. 
The average number of candidates per event 
before and after event selection 
are 10 and 1.1, respectively. 

Efficient track and photon 
quality requirements have been designed 
to minimize systematic uncertainties. 
This includes selecting only those photons 
that are detected in the region of the calorimeter 
where the resolutions are well modeled. 
Kaon candidates are required to have measured $dE/dx$ 
within $\pm2.5$ standard deviations ($\sigma$) 
of the expected energy loss. 
Pairs of photons combined to form the $\pi^0$ 
candidates are required to have masses within 
$-3.5\sigma$ and $+2.5\sigma$ 
($\sigma \approx 6$ ${\rm MeV/c^2}$) 
of the $\pi^0$ mass\cite{PDG}. 
To improve mass resolution for parent 
particles, the $\pi^0$ candidates 
are kinematically fit to this mass. 
To suppress combinatorial background, 
soft photons from the $D^{*0} \to D^0\gamma$ 
decays are required to have energies above 
200 MeV. 
This selection is 50\% efficient. 
The invariant mass of the $D^0$ candidates 
is required to be within 
$\pm2.5\sigma$ ($\sigma \approx  8.0 {\rm~MeV/c^2}$), 
$\pm2.0\sigma$ ($\sigma \approx 15.0 {\rm~MeV/c^2}$) 
and 
$\pm1.5\sigma$ ($\sigma \approx 7.5 {\rm~MeV/c^2}$) 
of the $D^0$ mass of $1.8646 {\rm ~GeV/c^2}$ 
in final states with one, two and three pions, 
respectively. 
The $D^{*0}$--$D^0$ mass difference 
$\delta M$ is required to be within 
$\pm2.0\sigma$ 
of $142.1 {\rm ~MeV/c^2}$\cite{PDG} 
($\sigma \approx 1.0 {\rm~and~} 5.0 {\rm~MeV/c^2}$ 
for the $\pi^0$ and $\gamma$ decays of the $D^{*0}$, respectively). 
To select $D^0 \to K^-\pi^+\pi^0$ candidates 
we require the $K^-\pi^0$ and $\pi^+\pi^0$ 
invariant masses to be consistent 
with the resonant substructure of the $D^0$ decays\cite{PDG}. 
%The efficiency of this requirement is 90\%. 
Continuum data were used to optimize these criteria 
to suppress combinatorial backgrounds.   

The major sources of background 
are photons from 
initial state radiation 
and from $\pi^0$ decays 
both from continuum and $B\bar{B}$ events. 
To suppress the real $\pi^0$ background 
and to reduce the cross-feed between 
the $\pi^0$ and $\gamma$ reconstruction 
channels of the $D^{*0}$, 
we apply a $\pi^0$ veto to the photons 
from both the $D^{*0}$ decay and the $\bar{B^0}$ decay. 
This is done by rejecting photons 
that, when combined with another photon candidate, 
form $\pi^0$ candidates 
within $-4.5\sigma$ and $+3.5\sigma$ 
of the $\pi^0$ mass. 
To suppress the remaining continuum background, 
we use a Fisher discriminant technique\cite{FISHER}. 
This discriminant is a linear combination 
of three angles and nine event shape variables. 
The first angle is between 
the $\bar{B^0}$ candidate momentum 
and the $e^+e^-$ collision (``beam'') axis. 
The second is the angle between 
the beam axis and 
the direction of the $\bar{B^0}$ candidate thrust axis. 
The third is the angle between 
the thrust axis of the $\bar{B^0}$ candidate 
and the thrust axis of the rest of the event. 
The nine event shape variables are the 
amount of energy detected in $10^\circ$ 
cones around the direction of the signal photon from the $\bar{B^0}$ decay. 
The Fisher discriminant coefficients 
are optimized to maximize the separation 
between continuum events that are jetlike 
and $B\bar{B}$ events that are 
spherical in shape at the $\Upsilon(4{\rm S})$ energy. 
This important selection criterion 
is optimized for each reconstruction channel 
separately using a combination of continuum data 
and simulated signal events, and 
has an efficiency between 40\% and 70\% 
depending on the reconstruction channel. 

We define the signal region in the 
two-dimensional plane of 
the beam-constrained $B$ mass 
$M(B) = \sqrt{E_{\rm beam}^2-p(B)^2}$ 
and 
the energy difference 
$\Delta E = E(B) - E_{\rm beam}$, 
where $E_{\rm beam}$ is the beam energy, 
$p(B)$ is the momentum of the $\bar{B^0}$ candidate 
and $E(B)$ is its detected energy. 
The signal region is defined by 
$M(B) > 5.275 {\rm ~GeV/c^2}$ 
and $|\Delta E| \le 100$ MeV. 
The $M(B)$ requirement is 1.5$\sigma$ 
below the actual $\bar{B^0}$ mass\cite{PDG} 
($\sigma \approx  2.8 {\rm~MeV/c^2}$). 
These criteria are optimized 
to suppress the cross-feed from $B$ decays 
to higher-multiplicity final states. 
The signal region selection is 78\% efficient. 

\begin{figure}[ht]
\centerline{
%clns	\psfig{figure=prl1.eps,height=3.60in,width=3.40in}
%clns	\psfig{figure=prl2.eps,height=3.60in,width=3.40in}
	\psfig{figure=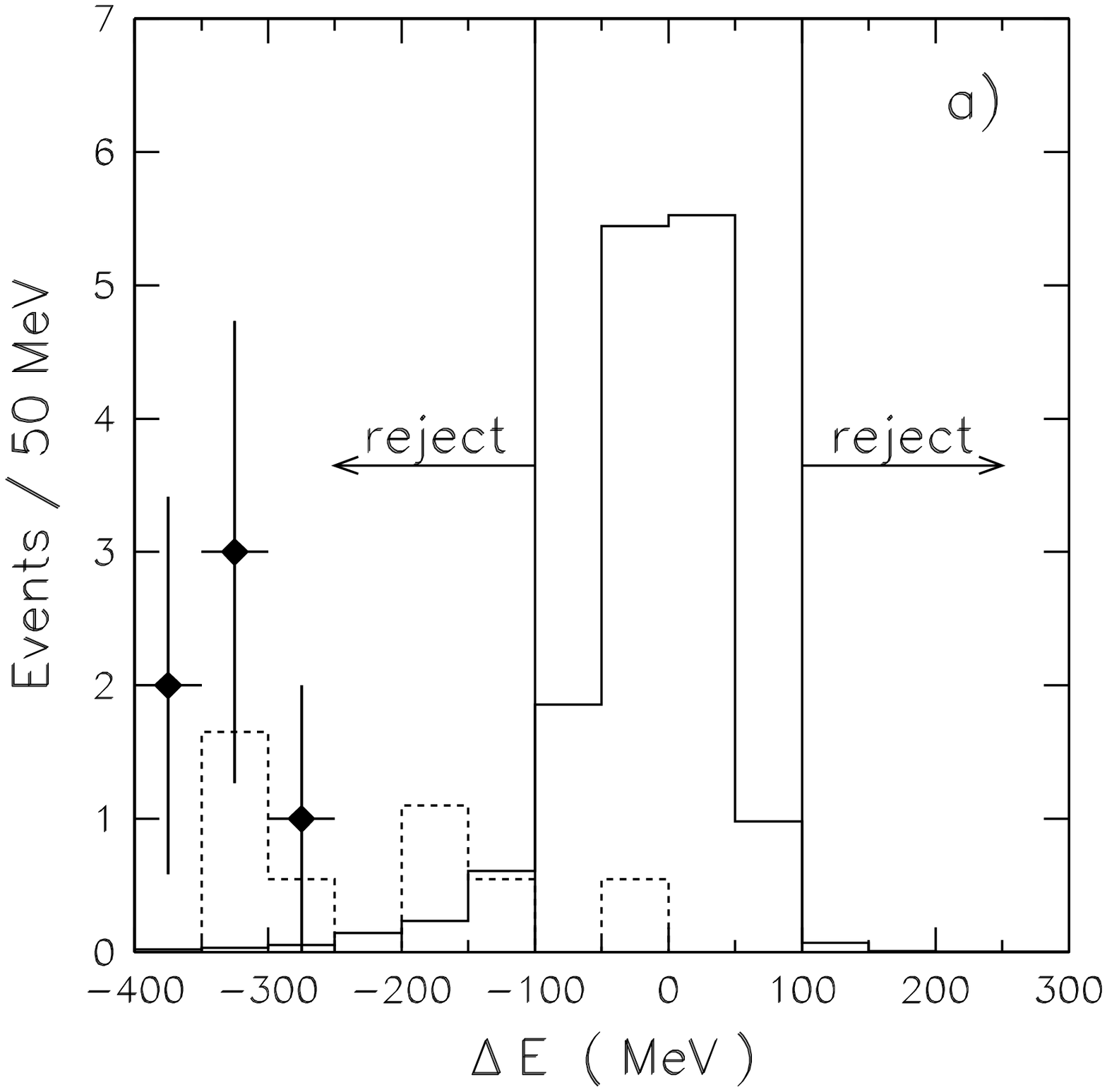,height=2.00in,width=1.80in}
	\psfig{figure=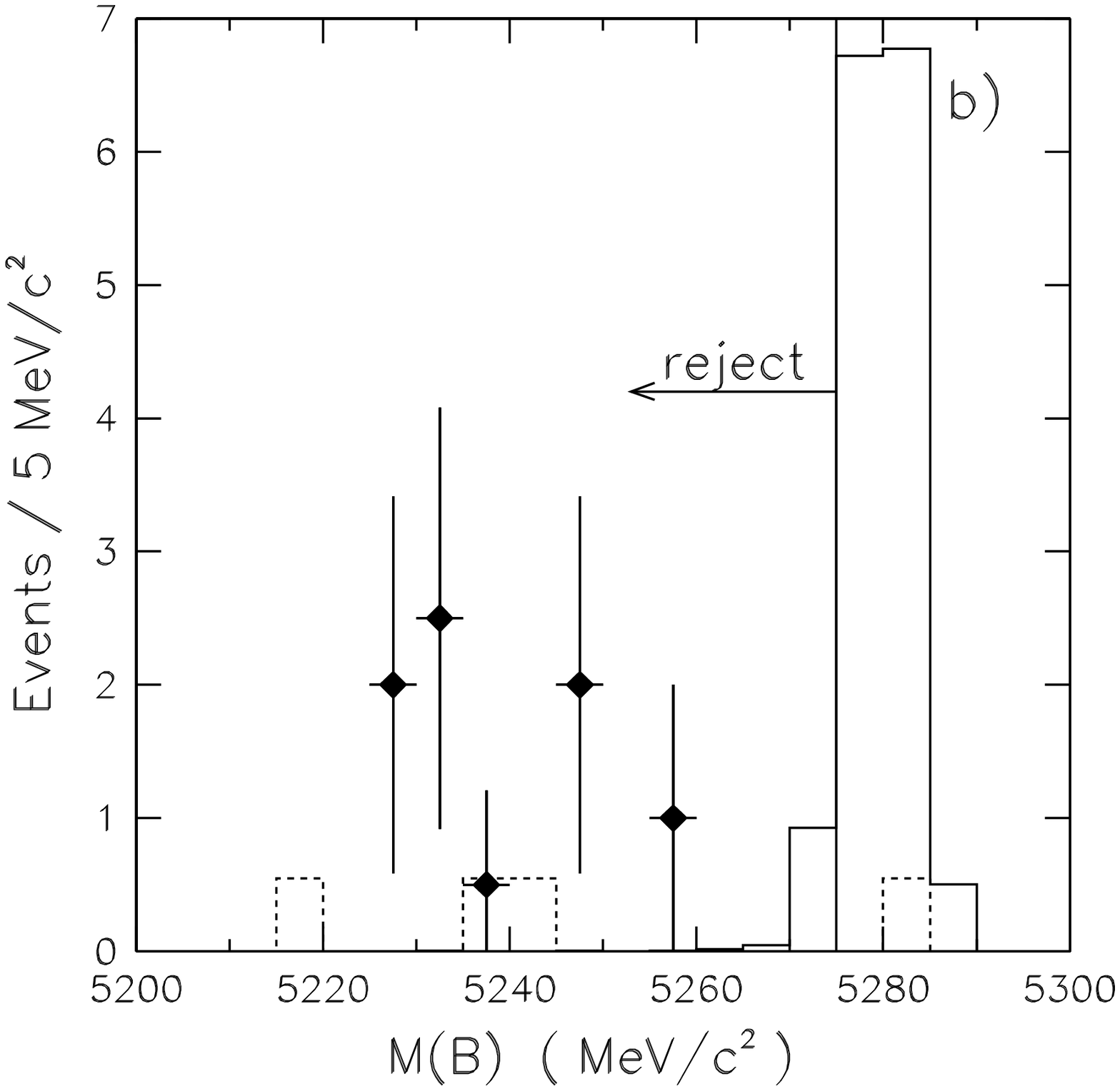,height=2.00in,width=1.80in}
%draft	\psfig{figure=prl1.eps,height=3.60in,width=3.40in}
%draft	\psfig{figure=prl2.eps,height=3.60in,width=3.40in}
           }
\caption{
a) $\Delta E$ and b) $M(B)$ projections of the signal region. 
Points show the $\Upsilon(4{\rm S})$ data, solid and dashed histograms 
show the predictions of the signal and $B\bar{B}$ simulations, 
respectively. 
The prediction of the $B\bar{B}$ simulation 
is normalized to the statistics of our data sample 
according to the $B$ branching fractions\protect\cite{PDG}. 
Simulated signal events are shown 
assuming that 
${\cal{B}}(\bar{B^0} \to D^{*0} \gamma) \approx 30 \times 10^{-5}$. 
The non-zero simulated $B\bar{B}$ contribution 
in the signal region 
is due to the decay $\bar{B^0} \to D^{*0} \pi^0$ 
assuming ${\cal{B}}(\bar{B^0} \to D^{*0} \pi^0) = 1.0 \times 10^{-4}$. 
Multiple candidates are weighted as described in the text. 
}
\label{figure}
\end{figure}

No events are found in the signal region. 
Projections onto the $\Delta E$ and $M(B)$ 
variables are shown in Fig.~\ref{figure}. 
On average we expect 0.5 continuum background events 
in the signal region. 
We estimate this number from continuum data by 
relaxing the event selection requirements. 
The contribution from the decay $\bar{B^0} \to D^{*0} \pi^0$ 
in the signal region is less than 0.9 events 
assuming 
${\cal{B}}(\bar{B^0} \to D^{*0} \pi^0) < 4.4 \times 10^{-4}$ at 90\% CL\cite{CLEO:color}. 
The theoretical predictions for this branching fraction 
are of the order of $10^{-4}$\cite{theory_6,theory_7}. 
%To estimate the cross-feed from the decay 
%$\bar{B^0} \to D^{*0} \pi^0$ we 
%have simulated a large dedicated sample of this decay.  
%Fig.~\ref{figure} shows 0.5 events from this channel in the 
%signal region from the simulated $B\bar{B}$ sample 
%matched to the data sample assuming 
%${\cal{B}}(\bar{B^0} \to D^{*0} \pi^0) = 1.0 \times 10^{-4}$. 
The contribution from all other known $B$ decays 
in the signal region is negligible. 
Six data events in the $\Delta E$ sideband 
are consistent with Monte Carlo expectations 
for the cross-feed from the decay $B^+ \to D^{*0}\rho^+$. 
This decay can produce 
$\bar{B^0} \to D^{*0} \gamma$ candidates 
with $\Delta E < -m_\pi$ when the $\pi^0$ decays 
asymmetrically and is emitted along the $\rho^+$ direction. 

To derive the upper limit we combine all 
six reconstruction channels. Efficiencies 
are weighted taking into account the 
branching fractions for the $D^{*0}$ and $D^0$ decays. 
The overall reconstruction efficiency is 2.3\%, 
where the major contributions are due to 
the exclusive reconstruction approach (30\%), 
the track and photon quality requirements (65\%), 
the $\delta M$ requirement (30\%) 
and 
the Fisher discriminant technique (58\%). 
To estimate the upper limit, we conservatively 
reduce reconstruction efficiency 
by its systematic error (18\%). 
The largest contributions to this error 
are due to the uncertainties 
in the track and photon reconstruction efficiencies (11\%), 
the $D^0$ branching fractions (9\%), 
Fisher discriminant (6\%) 
and 
the efficiencies of the requirements 
on the reconstructed masses of the $D^0$ (5\%) 
and $\bar{B^0}$ (5\%) candidates. 
To estimate the upper limit 
we assume ${\cal B}(\Upsilon(4{\rm S}) \to B^0\bar{B^0})={\cal
B}(\Upsilon(4{\rm S}) \to B^+B^-)=0.5$. 
The upper limit on the number of detected signal events 
is 2.3 at 90\% CL and corresponds to an upper limit 
on the branching fraction for the decay 
$\bar{B^0} \to D^{*0}\gamma$ 
of $5.0 \times 10^{-5}$ at 90\% CL. 
%according to Poisson statistics. 

We performed the first search 
for the decay $\bar{B^0} \to D^{*0} \gamma$ 
and set an upper limit on its branching fraction 
of $5.0 \times 10^{-5}$ at 90\% CL. 
Our non-observation is consistent with the absence of 
anomalous enhancements that could have overcome 
short-distance color suppression in the studied process. 
We confirm theoretical predictions 
that weak radiative $B$ decays are dominated 
by the short-distance $b \to s\gamma$ mechanism. 
Finally, our results should be useful for studies of radiative 
and color-suppressed processes with heavy quarks at future 
high statistics $B$ physics experiments. 
At these facilities the decay $\bar{B^0} \to D^{*0} \gamma$  
should be utilized to verify if the short-distance 
QCD radiative corrections are under firm theoretical control 
%numerically 
and, possibly, to search for new physical phenomena. 

We would like to thank 
A. Khodjamirian, 
P. Kim, 
R. Schindler, 
and 
A. Vainshtein 
for useful conversations. 
We gratefully acknowledge the effort of the CESR staff in providing us with
excellent luminosity and running conditions.
This work was supported by 
the National Science Foundation,
the U.S. Department of Energy,
the Research Corporation,
the Natural Sciences and Engineering Research Council of Canada, 
the A.P. Sloan Foundation, 
the Swiss National Science Foundation, 
and the Alexander von Humboldt Stiftung.  

%++++++++++++++++++++++++++++++++++++++++++++++++++++++++++++++++++++++++++++++++++++++++++++

\end{document}